\documentstyle[11pt,gh2001-asp,twoside,epsf]{article}
\def\edcomment#1{\iffalse\marginpar{\raggedright\sl#1\/}\else\relax\fi}
\reversemarginpar

\begin{document}
\title{Velocity Fields of Disk Galaxies}
\author{Peter J. Teuben}
\affil{Astronomy Department, University of Maryland, 
College Park, MD 20742, USA}

\begin{abstract}
Two dimensional velocity fields have been an important tool for nearly
30 years and are instrumental in understanding galactic mass
distributions and deviations from an ideal galactic disk.  Recently a
number of new instruments have started to produce more detailed
velocity fields of the disks and nuclear regions of galaxies. This
paper summarizes some of the underlying techniques for constructing
velocity fields and deriving rotation curves.  It also urges to
simulate observations from the data-cube stage to reject subtle biases
in derived quantities such as rotation curves.
\end{abstract}

\section{Introduction}
Two dimensional radial velocity fields of galactic disks are now
routinely derived from 
an ever increasing number of 
optical and radio
emission and absorption lines (HI, CO, H$\alpha$, [NII], ...) using a variety of
instruments (Radio and Fabry-Perot interferometers, integral field
spectrographs (IFS)).  These data are normally obtained as three
dimensional data-cubes, from which by either fitting some functional form to
the spectral line, or a moment analysis, a velocity field is derived. Such
two dimensional velocity fields are then fitted with a 
rotation curve\index{rotation curve}
by assuming circular rotation (Begeman 1989).
Analysis of rotation curves and comparison with models and
observations have led to the realization that dark matter in the
outer parts of galaxies must be the dominant gravitational force.
In addition the question of the contribution of the
stellar disk to that of the dark matter
(van Albada \& Sancisi 1986) and
the validity of Cold Dark Matter (de Blok et al. 2002) 
are both derived from simple
1-dimensional rotation curves. The process of
deriving a 1-dimensional rotation curves from a 3-dimensional data-cube 
is subject to many observational, instrumental and physical
biases, which we review here. Older reviews range
from the classic van der Kruit \& Allen (1978) paper
to last year's review by Sofue \& Rubin (2001). A number of 
(often Dutch\footnote{may contain a slight author bias})
PhD. theses also discuss many basic aspects
(Bosma 1978, Begeman 1987, 
Broeils 1992, de Blok 1997, Swaters 1999, 
Wong 2000) and although
sometimes hard to retrieve, are worth reading. In this paper
we will start from rotation curves, construct velocity fields, and
conclude by looking at how velocity fields are constructed from
data cubes.

%

\section{Rotating Galactic Disks}

Our ideal mathematical galactic disk is infinitesimally thin, 
and has material
rotating on circular orbits of negligible velocity dispersion.
Inclined at an angle $i$ to the sky plane, and
a major axis aligned with the $x-$axis for convenience,
the observed radial velocity is given by
\begin{equation}
        V(x,y) = V_{sys} + 
                 V_{rot}(R) \cos{\theta} \sin{i} 
                 + V_{exp}(R) \sin{\theta} \sin{i}      
\label{eq:vxy}
\end{equation}
where $(R,\theta)$ are polar coordinates measured in the plane of the
galaxy, and $(x,y)$ the cartesian coordinates on the plane of the sky.
By convention, position angles are measured counter clockwise from the
receding side, $V > V_{sys}$, of the galaxy major axis.  For later discussions
we add an expansion term, $ V_{exp}$; for circular orbits
this term will be 0 of course.

\begin{figure}[htbp]
\plotone{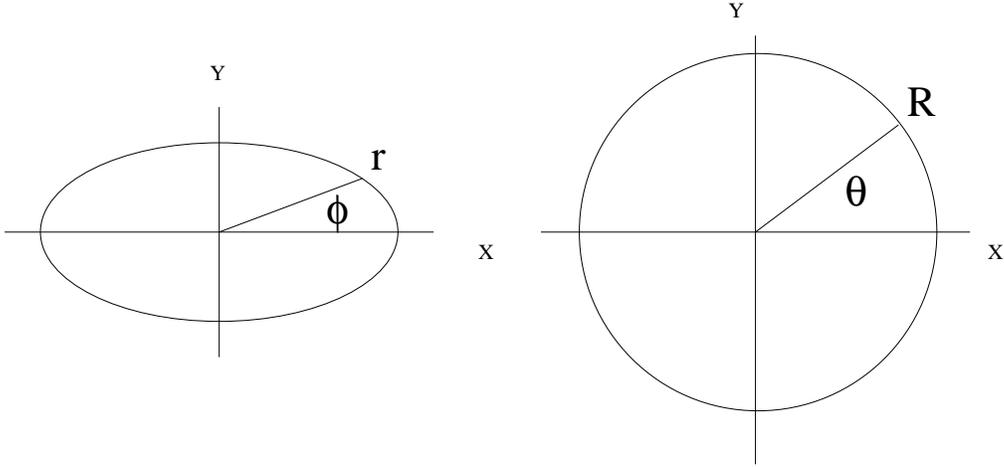}
\caption{Geometry of a galactic disk. On the left side as seen
projected on the sky plane, on the right as seen from above
the galactic disk, or $i=0$.}
\label{f:geom}
\end{figure}
The relationship between the sky and galaxy plane is given by:
\begin{equation}
        \tan{\theta} = { \tan{\phi} \over \cos{i}}, ~~~~~~
        R = r { \cos{\phi} \over \cos{\theta}}     \label{eq:geom}
\end{equation}

\subsection{Linear rotation curve (solid body)}

The centers of galaxies have often been assumed to have a solid
body (density $\rho(R) = constant$) 
rotation curve\footnote{Any steeper sloped density $\rho(R) \propto R^{-p}$ will
have a rotation curve $V(r) \propto R^{1-p/2}$ with infinite slope at the
center, but beam smearing will generally produce a linear rotation curve.}
\begin{equation}
        V(R) = \Omega R
\end{equation}
for which the velocity field (eq. (1)) is given by
\begin{equation}
        V(x,y) = \Omega x \sin{i}
\end{equation}
which means equi-velocity contours are lines parallel to the $y-$axis
(see Fig. 2a)
For beam size $B$ and tracer velocity dispersion $\sigma_g$ the observed
velocity dispersion will be
\begin{equation}
\sigma = \sqrt{ (B \Omega \sin{i})^2 + \sigma_g^2}
\end{equation}
but note that the mean velocity will not be affected by beam smearing
(assuming uniform surface density).

\subsection{Flat rotation curve (isothermal body)}

In the case of a flat rotation curve (density $\rho(R) \propto 1/R^2$)
\begin{equation}
        V(R) = V_0
\end{equation}
the velocity field is given by
\begin{equation}
        V(x,y) =  V_0  \sin{i} \cos{\theta} 
                = V_0 { {\sin{i} \cos{i}} \over \sqrt{\cos^2{i}+\tan^2{\phi}} }
\end{equation}
which means equi-velocity contours are now radial lines going through the center
(see Fig. 2b).
All along the major axis ($\phi=0,180$) the maximum amplitude $V_0 \sin{i} $ is observed,
and any beam smearing will now only include material
with lower radial velocities, and thus introduce
a bias against the maximum rotation speed.
For a beam size $B$ this lowest observed radial velocity is approximately
given by
\begin{equation}
V_{min} =   {   { R \cos i } \over  {  \sqrt{ { (R \cos i)^2} + B^2}  }} V_0
\end{equation}


\begin{figure}[htbp]
\plottwo{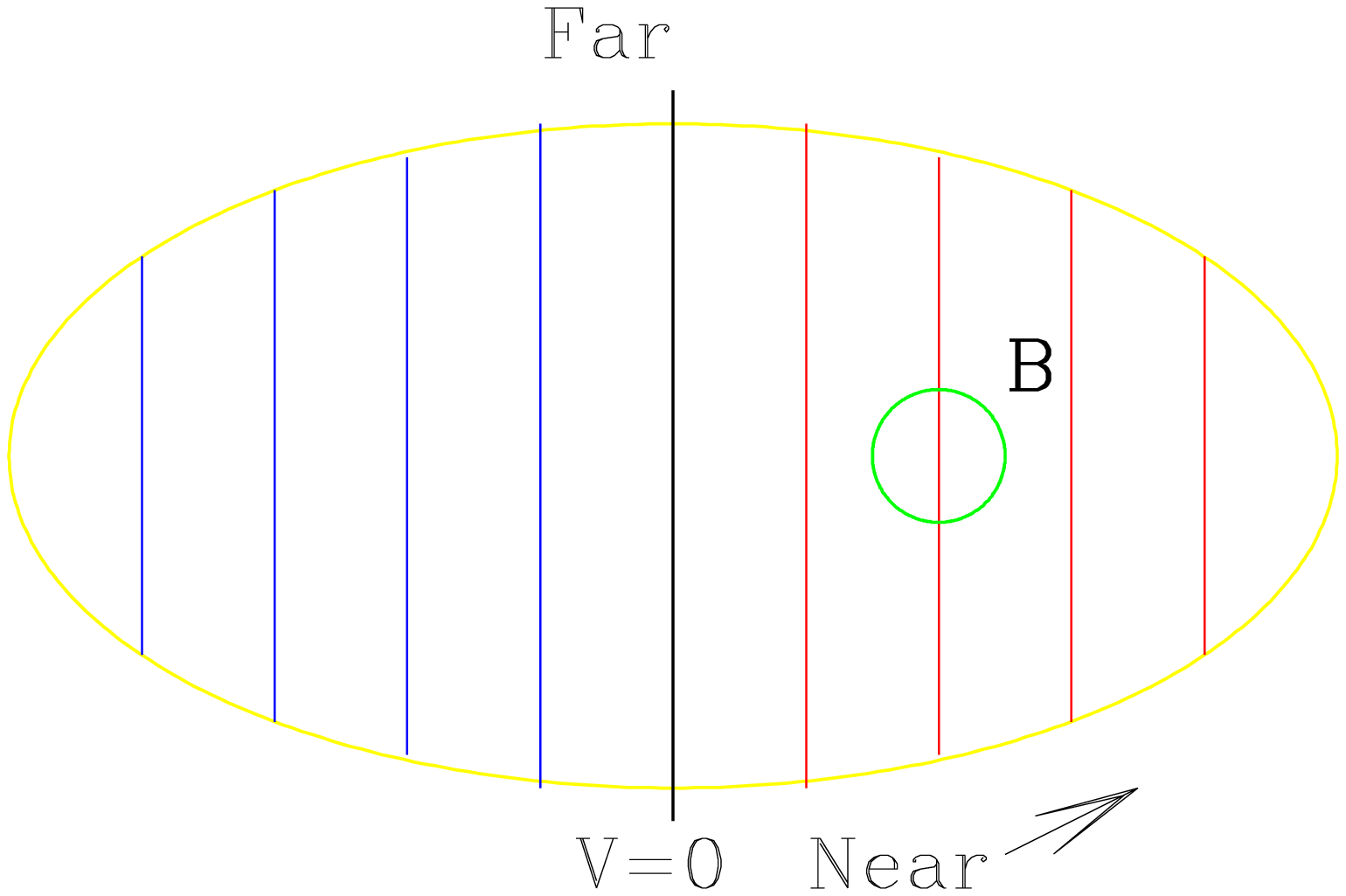}{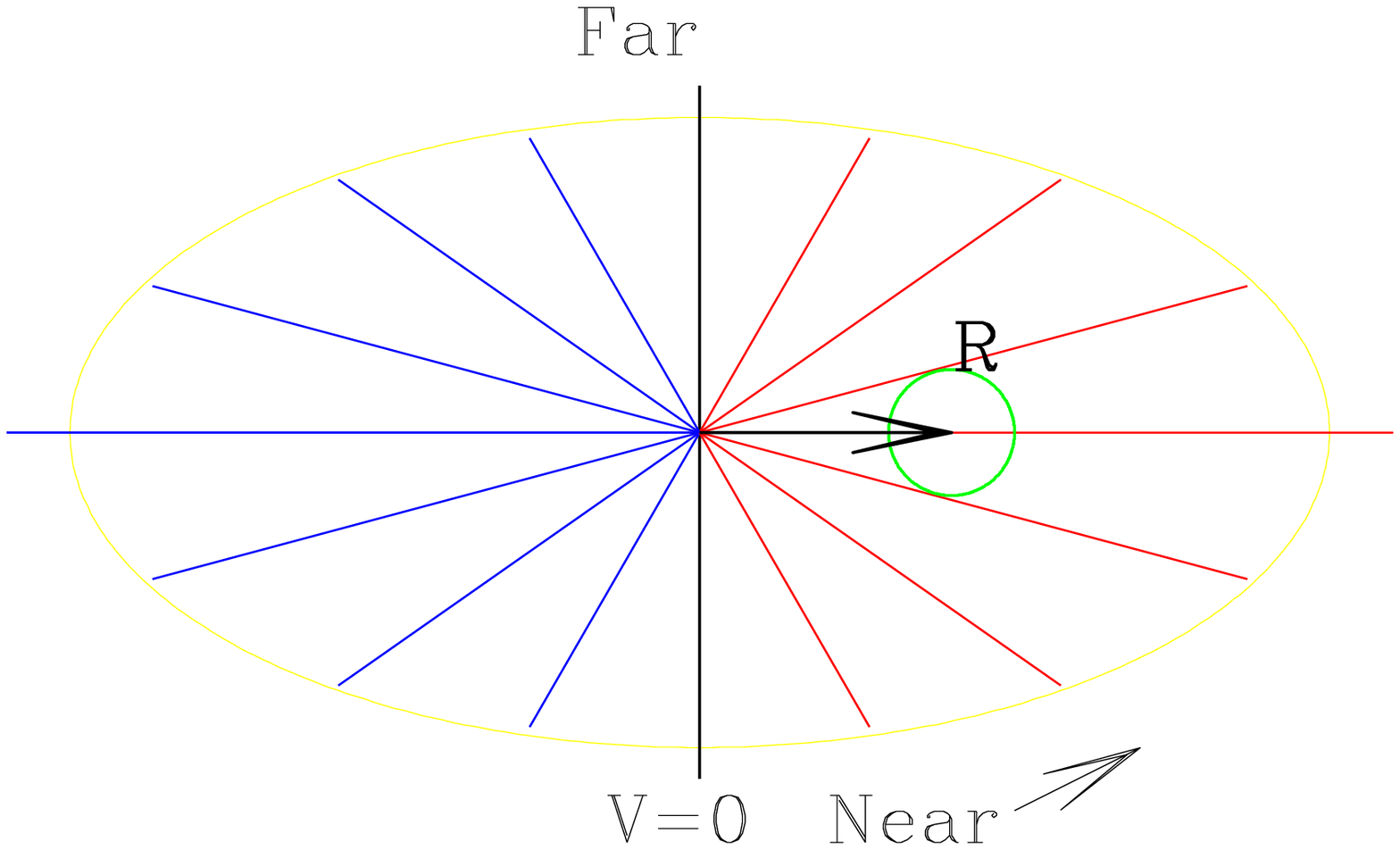}
\caption{Velocity fields of a linear (left) and flat (right) rotation curve. 
The receding side of the galaxy, $V > 0$, is on the right side of the galaxy).
Resolution beams of size $B$ have been sketched in as circles. Galaxies
are mostly a combination of these two: linear in the center, flat in the
outer parts, cf. Figure 4a.
}
\label{f:simple}
\end{figure}

An interesting property of velocity fields of projected 
circular orbits is that 
the kinematic major and minor axis, i.e. the lines
$V=0$ and maximum velocity gradient $V=min \rightarrow max$ 
are perpendicular to each other, and
these are also aligned with the morphological major and minor axis.
See also Figure 2.

\subsection{Deviations from the Ideal Galactic Disk}


Here is a list of some of the known deviations from an ideal galactic disk:

\begin{itemize}

\item The disk is not flat. For example warps (e.g. Bosma 1981)
or corrugations (Edelsohn \& Elmegreen 1997) are known forms of deviations
from perfectly flat disks. For warps the velocity field in each
annulus is still that of circular orbits, ignoring any precession
of the warp.

\item The disk is lopsided (m=1 mode) and will not support circular
orbits (see e.g. Noordermeer et al. 2002).

\item The disk is not axisymmetric (e.g. m=2 mode) and thus non-circular orbits 
($x_1,x_2$ orbits in bars, oval distortions, as well as non-spherical dark halos)
will be present. Best known examples are barred galaxies, where
deviations can easily reach 50-150 km/s (see e.g. Regan, Sheth, \& Vogel 1999).

\item Disks have spiral arms, i.e an m=2 mode with changing position angle.
Deviations from circular orbits can easily reach 10-50 km/s.
Spiral arms can also have associated features, such as spurs, that cause
higher order harmonics to be present in the velocity field. 
Recent simulations of 
spurs seem to create deviations or order 5 km/s (Kim \& Ostriker 2002).

\item Disks have a finite thickness, and are known to become flaring in
the outer parts. 
This will be important at high inclinations and near
the center where integrations along the line of sight become long.
They will widen the emission line profiles, but do not
necessarily remain Gaussian. See also Olling \& Merrifield (2000).

\item There can be radial outflows or inflows, often implied in the centers of
galaxies (but see also Fraternali et al. 2001, Schinnerer et al. 2000).

\item Turbulence in the ISM will add a mostly Gaussian component to the
line profiles.

\item Asymmetric drift corrections, when random motions provide
significant dynamical support (cf. Meurer et al. 1996) will need
to be applied to derive the correct dynamically derived mass
distribution:

\begin{equation}
 V_c^2 \simeq V_{rot}^2 -R \sigma_V^2 {{\partial \ln{\Sigma \sigma_V^2}} \over {\partial R}}
\end{equation}

\item Beam smearing, especially with high inclinations, 
will generally result in non-Gaussian profiles. 
Effects on the rotation curve can be 10-50 km/s if not corrected.

\item If the tracer is not uniformly distributed across the beam, there
will obviously be a bias in the derived velocity.
This is especially important where large velocity gradients are
present, such as near the galactic center (see also Figure 6). 
This will also give non-Gaussian profiles.

\item The bulge or nuclear potential can cause warps in the inner 
parts (see e.g. Schinnerer et al. 2000, where non-circular streaming also
appears to be present)


\item If the tracer is not optically thin, or if there is dust along the line of
sight, velocities will come out skewed. For given model distributions, reasonable
corrections can be made (see e.g. Baes \& DeJonghe 2001). For disk galaxies
this effect can be particularly important in 
the centers of galaxies, and in highly inclined galaxies.

\end{itemize}



\subsection{Elliptical orbits}

\index{elliptical orbits}
To first order
elliptical orbits will have a tangential and radial velocity 
\begin{equation}
        V_{rot} = V_0 + \epsilon  \cos{2\theta}, ~~~~~~
        V_{exp} = \epsilon  \sin{2\theta}
\end{equation}
with $\epsilon > 0$ for orbits oriented perpendicular to the major axis, and
$\epsilon < 0$ for orbits oriented along the major axis. Combining (1) and (10)
then gives:
\begin{equation}
        V(x,y) = V_{sys} + 
                ( V_0 + \epsilon) \cos{\theta} \sin{i}
\end{equation}
which is a signature of normal circular orbits.
In such degenerate cases it is not possible to detect elliptical
streaming  (see also Long 1991), but will show either a larger or lower
rotation curve, depending on the orientation of the ``bar'' w.r.t.
the galaxy major axis.

The more general case, with an ellipticity at an arbitrary angle
\begin{equation}
        V_{tan} = V_0 + \epsilon  \cos{2(\theta-\theta_0)}, ~~~~~~
        V_{rad} = \epsilon  \sin{2(\theta-\theta_0)}
\end{equation}
will result in a velocity field in which the kinematic major and minor
axis are not perpendicular anymore. Realistic orbits (and gas flow)
deviates considerably from simple ellipses (Athannassoula 1992),
but the general picture remains.

%
%

A Fourier analysis of the observed velocity field
(see e.g. Teuben 1991, Schoenmakers et al. 1997)
\begin{equation}
        V(x,y) = c_0 + \Sigma_m{(c_m \cos{m\phi} + s_m \sin{m\phi})}
\end{equation}
has also led to considerable insight in velocity fields of non-axisymmetric
potentials.
Spectral analysis of periodic orbits (Binney \& Spergel, 1982) provides
a natural translation from the periodic orbits to the 
Fourier components of the velocity field.

A simple inversion from a velocity field to a mass model for non-axisymmetric
mass distributions does not exist yet.
Sanders \& Tubbs (1980) approached this by searching 
for a least squares solution between
the observed and model gas flow velocity field, and used this to find the best
parameterized barred galaxy model description for NGC 5383.\index{object, NGC 5383}
Weiner et al. (2001) used this approach to break the degeneracy that normally
exists in decomposing an observed rotation curve into a dark matter
and an M/L converted disk component (``maximum disk hypothesis'').
\index{barred galaxies}

\section{Rotation Curve: velocity field fitting}

To derive a rotation curve from a velocity field, various approaches are possible:
Especially if the disk is suspected to have a warp,
the disk is divided in annuli, within which
the rotation speed and geometrical parameters can be fitted (see e.g. Begeman 1987).
One can also keep the geometry
for the whole disk fixed, and fit a single shape to the rotation curve
(see e.g. van Moorsel \& Wells 1985).
This will generally result in a better geometric definition of the disk,
but residual velocity fields should still be
investigated to confirm the absence of any systematic effects. Notable
problems are due to the product of $V_{rot} \sin{i}$ in eq.(1),
which makes it impossible
to determine a kinematic inclination for linear rotation curves,
and very hard for low values of the inclination.


The {\tt rotcur} program in GIPSY/NEMO fits the following function

\begin{equation}
        V(x,y) = V_{sys} + V_{rot} \cos{\theta} \sin{i}
                + V_{exp} \sin{\theta} \sin{i}
\end{equation}
with now in full glory:
\begin{equation}
        \cos{\theta} = {{ -(x-X_0) \sin{\phi_0} + (y-Y_0) \cos{\phi_0} }
                        \over
                        {r}}
\end{equation}
and
\begin{equation}
        \sin{\theta} = {{ -(x-X_0) \cos{\phi_0} - (y-Y_0) \sin{\phi_0} }
                        \over
                        {r \cos{i}}}
\end{equation}
for each ring. For circular orbits
there are 6 free parameters to each ring:
systemic velocity $V_{sys}$,  
rotation velocity $V_{rot}$, 
inclination $i$, 
position angle of the receding side of the galaxy $\phi_0$, 
and rotation center $X_0$, and $Y_0$. Alternatively
an expansion velocity, $V_{exp}$, term can be added to eq.(13), increasing
the number of free parameters of this non-linear fit to 7.  Also note
although each ring is fit and provide just a convenient geometrical
description, this has not included any real dynamical effects, such
as precession, of those rings in a true warped galaxy.
For elliptical streaming a phase shifted $2\theta$ component (cf. equation (11))
is added to $V_{rot}$ and $V_{exp}$, causing $\theta$ and $3\theta$ harmonics
in the radial velocity field. Attempting a tilted ring fit will take care
of the $\theta$ harmonic, leaving a characteristic 
$3\theta$ residual (see e.g. Teuben (1991) Figure 3).


\begin{figure}[t]
\plotone[79 306 560 654]{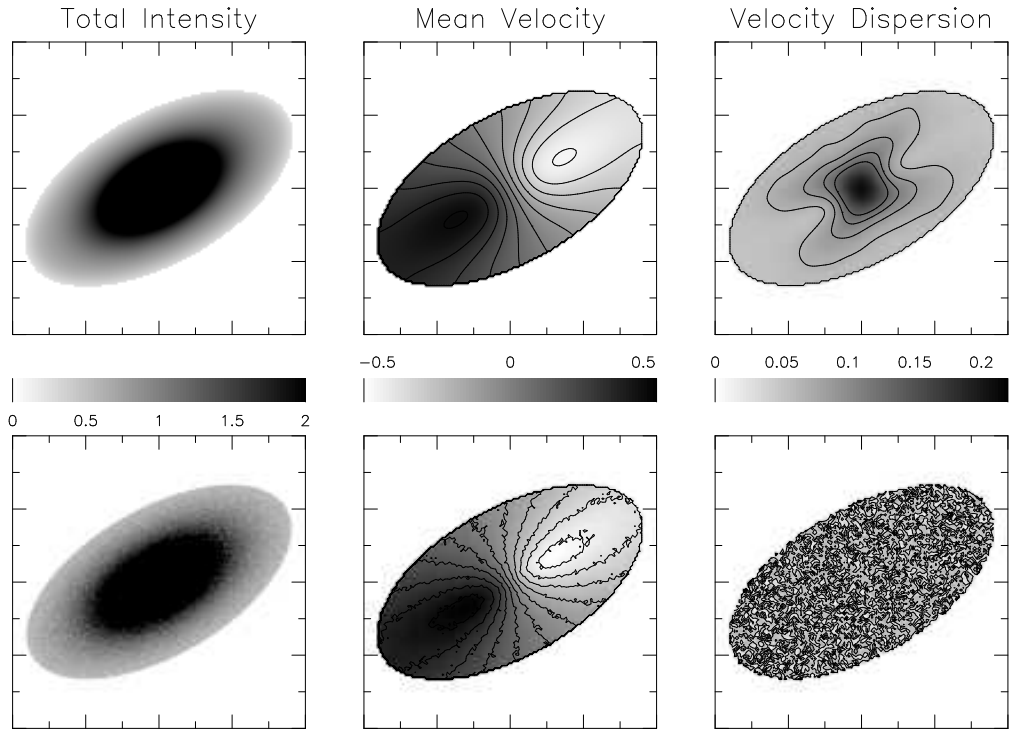}
\caption{
Total intensity (left), mean velocity (middle) and 
velocity dispersion (right) for a disk galaxy with beam smearing
(top) and without (bottom). Notice the characteristic H shaped
velocity dispersion map and the degraded dV/dR accross the center.
In this example the beam is 0.6 (each tickmark is 1.0) and
the velocity gradient was measured to decrease by 25\%, 
from 0.96 to 0.72.
}
\label{f:moments}
\end{figure}

Although these methods have been applied to numerous galaxies, observed
in the HI, H$\alpha$ and CO lines, recent integral field spectographs
are revolutionizing this field because of their high signal-to-noise.
Kinematic inclinations can be reliably measured as low as $20\deg$
(see e.g. contributions by  Andersen \& Bershady  in this volume).
Non-circular motions (bars, spiral arms) 
now become the limiting factor in
deriving geometric parameters from a velocity field, making it
possible to test subtle differences in models.


The tilted-ring fit has 6 free parameters per ring, which makes
finding a unique solution hard.
In practice one often takes a number of steps to bring the number
of parameters down. For example, one determines the center
($V_{sys}$, $X_0$, and $Y_0$) either from a number of broad rings
where $i$, $\phi_0$ are kept fixed at reasonable values (e.g. from
optical or IR photometry). Alternatively, since the velocity field
near the center can often be assumed to be that of a
linear rotation curve, one can also
determine $V_{sys}$ from a least squares planar fit to
$V(x,y) = V_{sys} + a (x-X_0) + b (y-Y_0)$ where now the
galactic center $(X_0,Y_0)$ is determined from optical or
IR photometry. Both methods should be compared,
to exclude systematic effects such as $m=1$ modes
(cf. Beauvais \& Bothun 1999, 2001).

Next one fixes $V_{sys}$, $X_0$, and $Y_0$, and fits
$V_{rot}$, $i$, and $\phi_0$ in a number of broad rings. 
The position angle is usually well determined, whereas
the scatter in inclination can be large (Begeman 1989). If there
is a clear trend in position angle and/or inclination,
a kinematic warp may be present
and may have to be corrected for (though elliptical orbits can
also mimic a changing inclination, as well as interesting
effects such as shown in Figure 4 and 5).

Additionally one can also fix $\phi_0$, and just fit
$i$ and $V_{rot}$,
again in a small number of broad rings.  This should bring down the
error in $i$, and a weighted average should result in a 
a good value for the inclination. 

Finally, after having checked for kinematic warping, 
one fixes (or trends) all geometric parameters
and just fits for $V_{rot}$ in smaller rings, ideally the width of a beam
and interlacing the radii of the rings by half a beam width to obtain a reasonably
sampled rotation curve.


\section{Data Analysis}

The most common technique to derive a velocity field from a data cube 
is a pixel-based profile analysis. Depending on the data quality and 
type of observations, a number of different techniques have been employed: 

\begin{itemize}
\item single gaussian (or Voigt) fit. Works well for good signal-to-noise data,
        and can also easily be extended when multiple components are present.
\item intensity averaged (``mean'') velocity, useful for low signal-to-noise data.
\item median velocity, a more robust alternative to mean velocity.
\item peak (or peak fit) velocity.
\item window (converging mean algorithm, see Bosma 1978)
\item largest velocity from a multiple component gaussian fit (Begeman 1989)
\item envelope tracing (Sofue \& Rubin 2001)
\item Gaussian-Hermite moments (van der Marel \& Franx 1993)
\item Fourier Quotient (Bender 1990)
\end{itemize}

When the bandwidth of the observations is large compared to that
of the spectral line, and to limit the influence of noise, 
most of these techniques benefit 
by applying some kind a mask over the data
where emission is believed to be present. This can be most
effectively done by using
a highly spatially (and perhaps spectrally) convolved datacube where the 
signal-to-noise is higher and defining the mask based on a simple 
cutoff value in this convolved cube.
Iterative schemes which result in maps with varying degree of
spatial resolution are also possible, but care has to
be given that the new data is not just convolved data from 
already observed neighboring points (Vogel et al. 1993).

For low signal-to-noise data one can also use the fact that emission
is on a ``wiggly sheet'' in the data cube, and interactively
define emission in a set of 
position-velocity\index{position-velocity diagram}
diagrams. Although
subjective and labor intensive, this last resort can significantly
improve the velocity field. An additional aid can be setting a 
liberal masking window in velocity space around the expected
radial velocity on an earlier iteration on the rotation curve.

\begin{figure}[htbp]
\plottwo{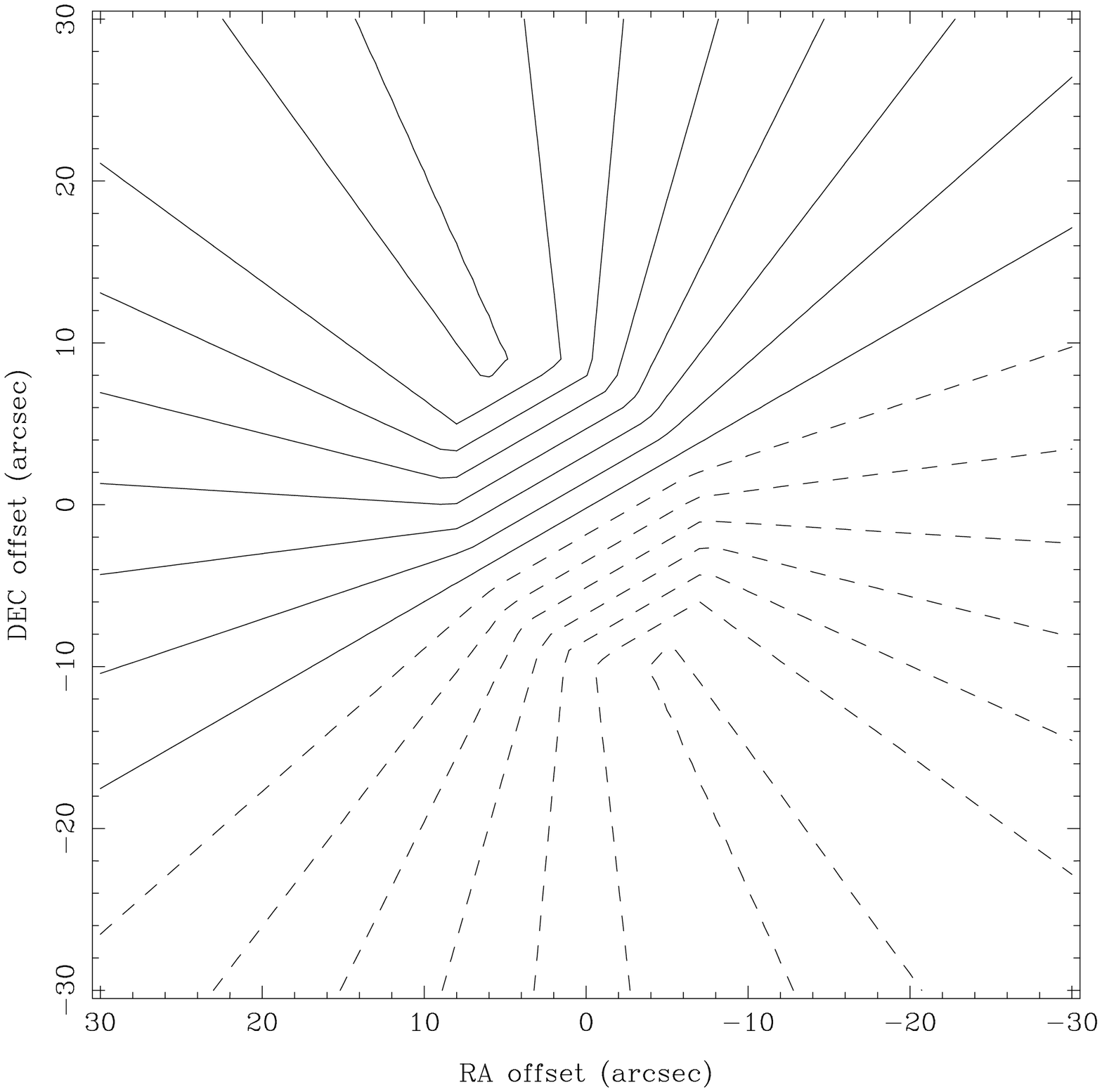}{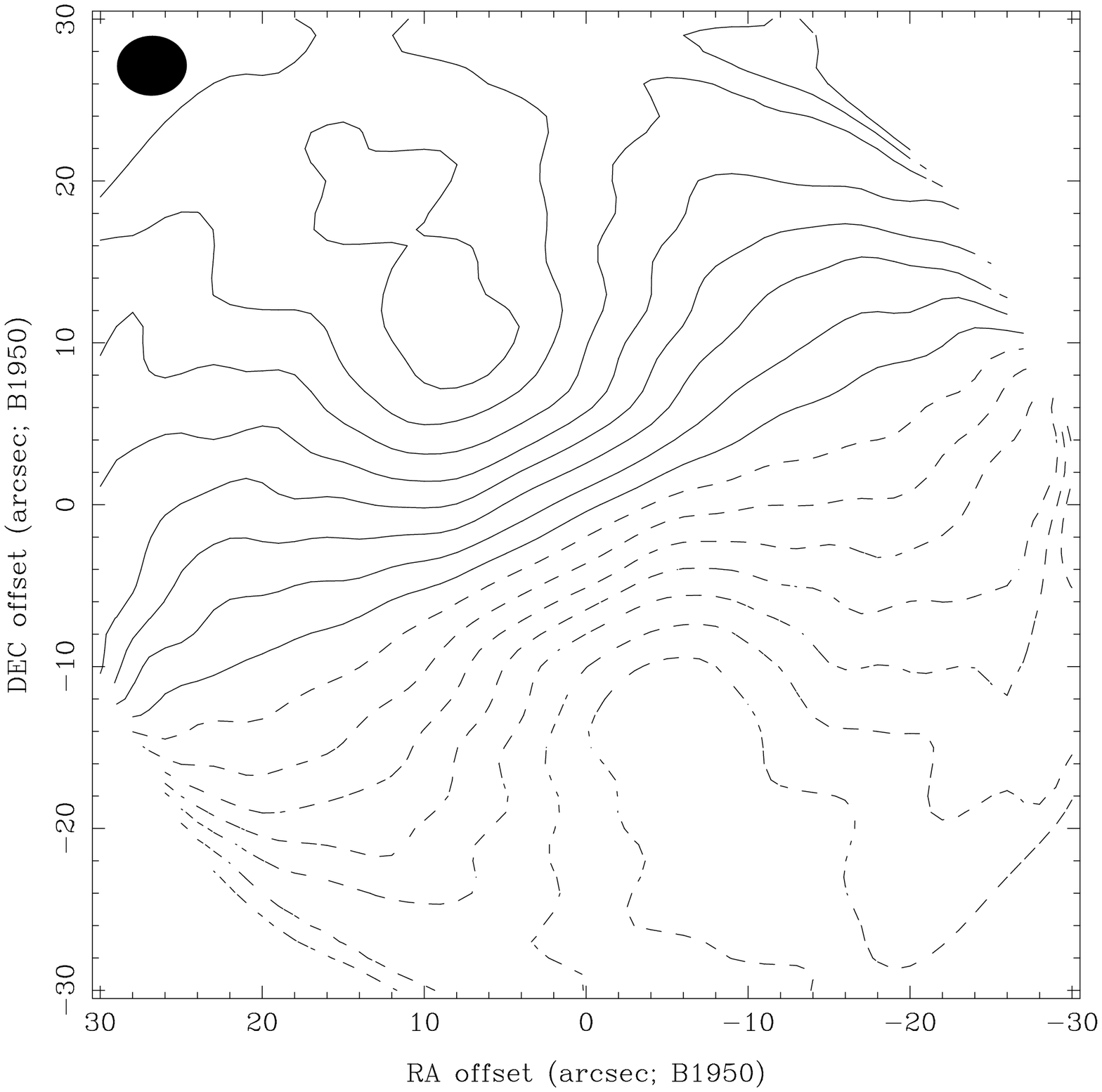}
\caption{Example of modeling: 
comparing a moment-1 derived velocity field (right panel,
with its synthesized beam in the top left corner) of a
radio interferometric observation with the input model (left). 
Contours on the approaching side of the galaxy are dashed. 
See Figure 5 for the corresponding tilted ring fit,
showing that special care may be needed for such rare cases
of near uniform density.
}
\label{f:velfie}
\end{figure}

For highly inclined galaxies (of which our own Galaxy is a special case)
some type of envelope tracing\index{envelope tracing}
technique has been successful in retrieving a rotation curve 
(Sofue \& Rubin 2001, see also Shane \& Bieger-Smith 1966).
If the ISM can be resolved, such as is the
case in our galaxy, a gaussian decomposition selecting the highest velocity
component can also be a  successful method.
Kregel (2002) recently introduces the ``union peel'' method for edge-on 
galaxies.

A full cube fit is the final resort when various geometric and beam smearing
effects all take too much effect. An input model then predicts the resulting
datacube and iteratively corrections are made to the model until the observational
cube matches the theoretical one (e.g. Swaters 1999)

\subsection{Modeling}

With current computing powers detailed simulations of the observations
show (Teuben et al., in prep.) that a datacube can have subtle biases,
which in turn \index{velocity field, modeling}
adds a bias to
the velocity field, which will then result in a biased rotation curve!
Consider the case of an interferometric observation of a typical galactic disk.
Channel maps around the systemic velocity have features aligned along the minor
axis, whereas near both of the extreme velocity channels tend to have features 
predominantly aligned along the major
axis. Depending on such details as the distribution of visibilities in the 
$U-V$-plane and adding short spacings to the visibilities, deconvolution will
result in differently recovered features in these channel maps. This is 
illustrated
in Figure 4 for a model and simulated velocity field.
The input model consisted of a gaussian disk with a FWHM size of 50\arcsec\ with
a linearly rising rotation curve to $R=10$\arcsec\ and $V=100$ km/s, and flat at larger 
radii. A single track with 10 antennae in the BIMA C-array has been used
to simulate the observation of this model disk. After mapping and 
(CLEAN) deconvolution, an ``observed'' velocity field was constructed by
computing an intensity weighted mean velocity by clipping signal
above the noise level. After this {\tt rotcur} was used to retrieve
the rotation curve, and geometric parameters. Only the systemic velocity
and rotation center were fixed, whereas the remaining 3 parameters were
fitted.
Figure 5 shows the derived parameters from such 
a tilted ring fit.
Notice that the position angle has been underestimated, and that the disk
appears to be slightly warped in nature.

\begin{figure}[htbp]
\plotone[40 170 570 700]{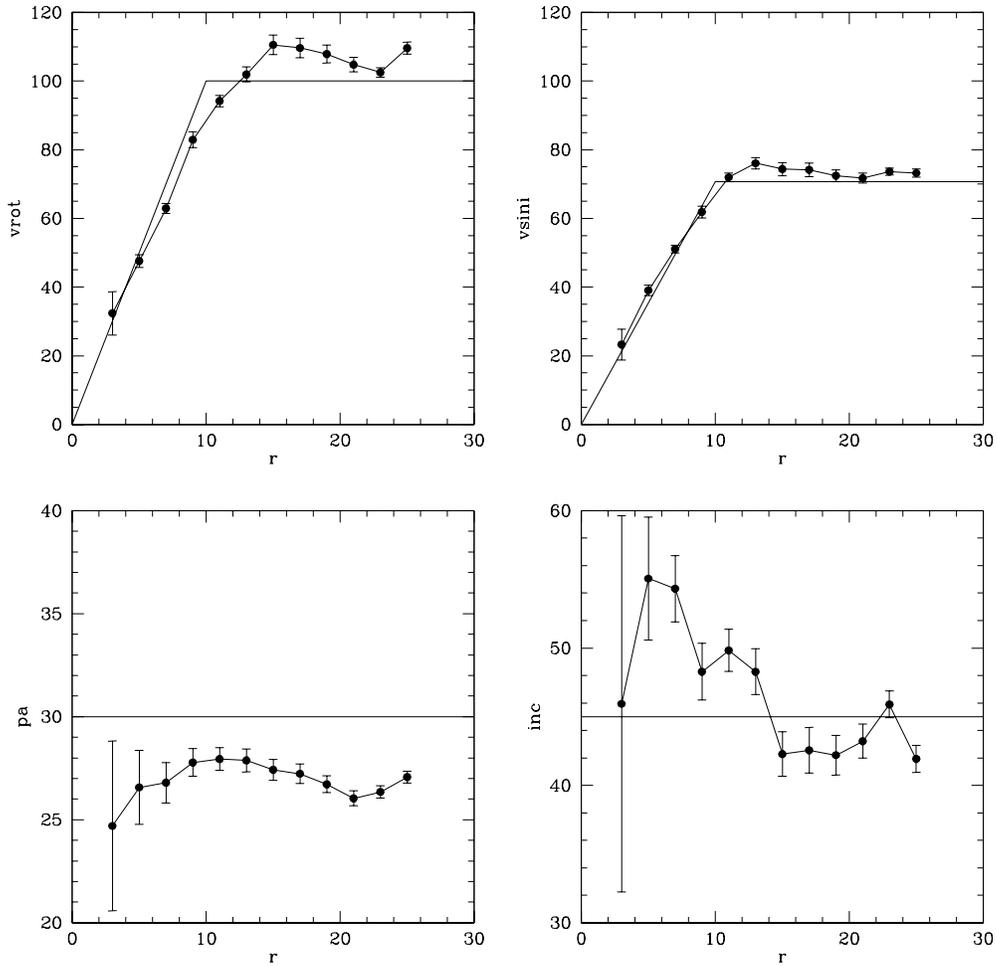}
\caption{Tilted ring rotation curve fits from simulated observations.
Top left: rotation curve. Top right: 
$V \sin{i}$.
Bottom left: position angle. Bottom right: inclination. Solid lines are the
expected values from the input model. The center position and
systemic velocity were kept fixed at their expected values.
}
\label{f:rotcur}
\end{figure}

\subsection{Smoothing}

\index{beam smearing}\index{smoothing}
For a non-uniform distribution of material across the beam, there
can be a bias of the derived velocity, independent of the methodology.
Figure 6 shows some of this effect
by measuring the gradient of a solid body rotation curve and comparing it
to the input value. For decreasing density distributions the velocity will
be biased to lower values and thus decrease the measured velocity gradient.

\begin{figure}[htbp]
\plotone[30 160 570 430]{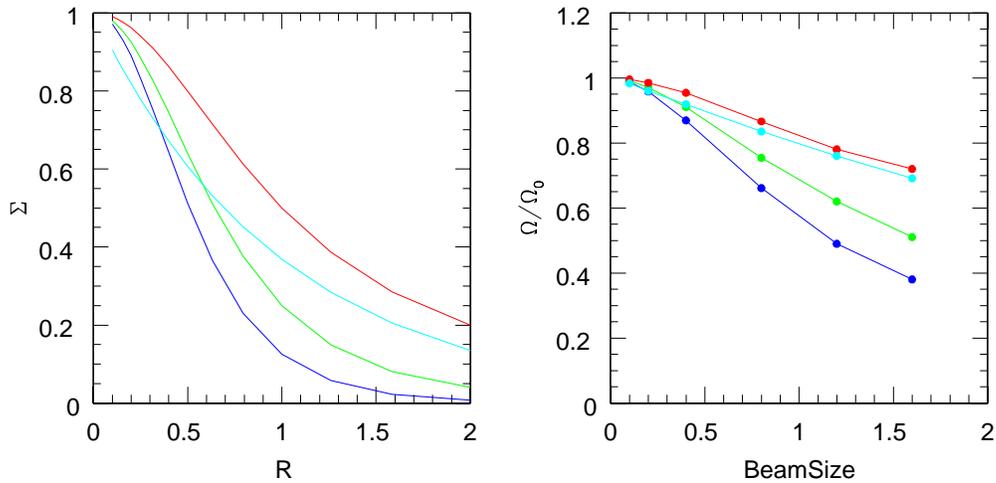}
\caption{
Different density distributions (left) as a function of radius
result in different measured dV/dR for a linear rotation curve 
as compared with the expected value (right). These were derived from
an intensity weighted mean velocity rotation curve.
}
\label{f:smoothing}
\end{figure}

\section{Conclusions}

Velocity fields of disk galaxies hardly ever show that of 
a perfectly rotating disk with circular orbits. Many deviations
have been identified, and can be readily derived by studying
residual velocity fields.
We have also shown some of the complexities that are involved when
extracting 1-dimensional rotation curves from 3-dimensional
datacubes. In extreme cases care has to be given to model
the observations, as they may introduce subtle biases in
the process of determining velocity fields and derived 
parameters.


\acknowledgements
I wish to thank the Guillermo Haro program for a scientifically stimulating
and socially pleasant atmosphere during the workshop and conference, and
Rosario Sanchez for keeping my cough under control.
I also wish to thank the BIMA SONG team for discussions and fabulous data.
Marc Verheijen is acknowledged for discussions around their IFS results.
This research was supported by NSF grant AST-9981289.

\end{document}